\documentclass[amsmath,amssymb]{nature_arXiv}

\usepackage{bm}
\usepackage{cite}
\usepackage{graphicx}

\title{Higgs transition from a magnetic Coulomb liquid to a ferromagnet in Yb$_2$Ti$_2$O$_7$}

\author{Lieh-Jeng Chang$^{1,2}$, Shigeki Onoda$^3$, Yixi Su$^4$, Ying-Jer Kao$^5$, Ku-Ding Tsuei$^6$, Yukio Yasui$^{7,8}$, Kazuhisa Kakurai$^2$ \& Martin Richard Lees$^9$.
}
\begin{document}

\maketitle

\begin{affiliations}
\item Department of Physics, National Cheng Kung University, Tainan 70101, Taiwan,
\item Quantum Beam Science Directorate, Japan Atomic Energy Agency, Tokai, Ibaraki 319-1195, Japan,
\item Condensed Matter Theory Laboratory, RIKEN, Wako, Saitama 351-0198, Japan,
\item J\"{u}lich Centre for Neutron Science JCNS-FRM II, Forschungszentrum J\"{u}lich GmbH, Outstation at FRM-II, Lichtenbergstrasse 1, D-85747 Garching, Germany,
\item Department of Physics and Center for Advanced Study in Theoretical Science, National Taiwan University, Taipei 10607, Taiwan,
\item National Synchrotron Radiation Research Center, Hsinchu 30076, Taiwan,
\item Department of Physics, Division of Material Science, Nagoya University, Furo-cho, Chikusa-ku, Nagoya 464-8602, Japan,
\item Department of Physics, Meiji University, Kawasaki 214-8571, Japan
and
\item Department of Physics, University of Warwick, Coventry CV4 7AL, United Kingdom.
\end{affiliations}

\begin{abstract}
In a class of frustrated magnets known as \textit{spin ice}, magnetic monopoles emerge as classical defects and interact via the magnetic Coulomb law. With quantum-mechanical interactions, these magnetic charges are carried by fractionalised bosonic quasi-particles, spinons, which can undergo Bose-Einstein condensation through a first-order transition via the Higgs mechanism. Here, we report evidence of a Higgs transition from a magnetic Coulomb liquid to a ferromagnet in single-crystal Yb$_2$Ti$_2$O$_7$. Polarised neutron-scattering experiments show that the diffuse $[111]$-rod scattering and pinch-point features which develop on cooling are suddenly suppressed below $T_{\mathrm{C}}\sim0.21$~K, where magnetic Bragg peaks and a full depolarisation of the neutron spins are observed with thermal hysteresis, indicating a first-order ferromagnetic transition. Our results are explained on the basis of a quantum spin-ice model, whose high-temperature phase is effectively described as a magnetic Coulomb liquid, while the ground state shows a nearly collinear ferromagnetism with gapped spin excitations.
\end{abstract}


\section*{Introduction}

Emergent magnetic monopoles\cite{Dirac:31} have recently been identified in many condensed-matter systems\cite{Hermele:04,Castelnovo:08,NSOMO:10,Qi:09}. In particular, recent experiments\cite{Sakakibara:03,Morris:09,Fennell:09,Giblin:11}  have shown that the low-temperature state of dipolar spin-ice\cite{Bramwell:01,MoessnerRamirez:06} materials, Ho$_2$Ti$_2$O$_7$ and Dy$_2$Ti$_2$O$_7$, can be effectively described as a classical magnetic Coulomb phase, which hosts these emergent magnetic monopoles obeying the magnetic Coulomb law\cite{Hermele:04,Castelnovo:08}. Theoretically, this has been understood from a classical model where $\langle111\rangle$ Ising moments, i.e., pseudospin-$1/2$, interact mainly through a magnetic dipolar interaction\cite{Bramwell:01}. This classical Coulomb phase is characterised by dipolar spin correlations with a power-law decay\cite{Isakov:04,Henley:05}. 

Quantum variants of spin ice\cite{Molavian:07,Onoda:10prl,Onoda:10} have also been found in the magnetic pyrochlore oxides RE$_2$TM$_2$O$_7$~(ref.~\citeonline{Gardner:10}) with the rare-earth (RE) ions Yb~(refs~\citeonline{Blote:69,Hodges:01,Hodges:02,Yasui:03,Gardner:04,Ross:09,Onoda:11,Balents:11}) and Pr~(refs~\citeonline{Zhou:08,Machida:10,Onoda:10prl,Onoda:10}) and transition-metal (TM) ions such as Ti and Ir, or Sn. In this quantum version of spin ice, the magnetic monopole charges are expected to be carried by fractionalised pseudospin-$1/2$ quantum-mechanical quasi-particles, spinons, that acquire kinetic energy by hopping via a pseudospin-flip exchange interaction, if they remain deconfined with an energy gap\cite{Savary:11}. Thus, these systems provide a unique playground to study the intriguing quantum dynamics of monopolar spinons. Quantum spin ice actually harbours the analogous ``quantum electrodynamics'', where the ``electric field'' and ``electric charge'' describe the Ising magnetic moment and the associated monopole charge, respectively\cite{Hermele:04}. In particular, Yb$_2$Ti$_2$O$_7$ provides a prototypical system described by an effective pseudospin-$1/2$ quantum spin-ice model\cite{Onoda:11,Balents:11} showing strong transverse quantum fluctuations of magnetic dipoles as a result of the following three characteristics: First, the crystal field structure of the Yb$^{3+}$ ion is characterised by a ground-state Kramers doublet well-separated from the first excited doublet\cite{Hodges:01}. Second, the Curie-Weiss temperature is weakly ferromagnetic with $\Theta_{\rm CW}\sim 0.53$~K~(ref.~\citeonline{Yasui:03}) as in dipolar spin ice\cite{Bramwell:01}. Third, the $g$-tensor has planar components, $g^\perp=4.18$, that are larger than the $\langle111\rangle$ Ising components $g^\parallel=1.77$~(ref.~\citeonline{Hodges:01}), enhancing the contributions to the magnetic neutron-scattering cross-sections from planar fluctuations.

An early specific-heat measurement on a powder sample of Yb$_2$Ti$_2$O$_7$ revealed a sharp singularity at 0.21~K, signalling a first-order phase transition\cite{Blote:69}. A previous neutron-scattering experiment, on a single crystal in which we have subsequently shown that the specific-heat anomaly is reproduced (Supplementary Figure~S1), found an almost collinear ferromagnetic structure with an ordered Yb$^{3+}$ magnetic moment of $1.1\mu_B$, a uniform magnetisation along $[100]$ and a slow relaxation of the spins over times of the order of two hours well below $T_{\mathrm{C}}\sim0.21$~K~(ref.~\citeonline{Yasui:03}). These results may indicate the formation of an energy gap in the spin excitations below a first-order phase transition to the ferromagnetic state. We have shown from specific-heat measurements (Supplementary Figure~S1) and extended X-ray absorption fine structure (EXAFS) experiments (Supplementary Figure~S2) on three different single crystals (see Methods), that the long-range ferromagnetic order in Yb$_2$Ti$_2$O$_7$ is quite sensitive to Yb deficiency. In particular, our analysis indicates that as the Yb deficiency increases, the specific-heat anomaly becomes significantly broader and the sharp anomaly is lost (Supplementary Figure~S1). This provides an understanding of the experimental results reported for different single-crystal and polycrystalline samples of Yb$_2$Ti$_2$O$_7$, which suggest magnetic correlations remain short-range and dynamic even below 0.21~K~(refs~\citeonline{Hodges:02,Gardner:04,Ross:09}). The variation in the experimental findings on different samples suggests the proximity of this material to a quantum phase transition between a quantum magnetic Coulomb phase (U(1) spin liquid)\cite{Hermele:04} and a magnetically ordered phase (Fig.~1~\textbf{a}). 
The magnetic ordering in this case can be understood as a Bose-Einstein condensation of monopolar spinons within the Higgs mechanism\cite{Fradkin:79,RK}, which gaps out all the soft spin excitations and takes a first-order character\cite{Savary:11, Fradkin:79,RK}, as in the case of normal-to-superconductor transitions\cite{Anderson:63}. This first-order quantum phase transition should extend to finite temperatures\cite{Banerjee:08}. Suppose that the transition temperature is much lower than an energy $2J$ required for thermally and dynamically exciting monopolar spinons and that the high-temperature phase does not suffer from a static and thus non-uniform proliferation of monopole charges that causes a spontaneous symmetry breaking. Then, the high-temperature phase may be described as a magnetic Coulomb liquid at temperatures lower than $2J$, as in dipolar spin ice. Furthermore, the low-temperature phase transition can be effectively described as a Higgs transition of magnetic monopoles, like that of electric monopoles in the classical case\cite{Powell}. The minimal requirement for the identification of this Higgs transition of magnetic monopoles is the observation of a first-order phase transition separating a high-temperature Coulomb phase that may be characterised by pinch-point features\cite{Fennell:09,Morris:09,Isakov:04,Henley:05} and a low-temperature magnetically ordered phase that exhibits finite planar components of pseudospins and gapped spin excitations.

In this letter we report evidence of a Higgs transition from a magnetic Coulomb liquid to a ferromagnet, from data which were collected using low-temperature polarised neutron-scattering experiments on a single crystal of Yb$_2$Ti$_2$O$_7$. We observe that well below $T_{\mathrm{C}}\approx 0.21$~K the incident polarised neutron beam is completely depolarised. This is ascribed to the formation of macroscopic ferromagnetic domains. The first-order nature of the phase transition is revealed via the observation of a thermal hysteresis around $T_{\mathrm{C}}$.  A nearly collinear ferromagnetic structure consistent with previous neutron-scattering experiments on the same single crystal\cite{Yasui:03} is identified theoretically using an effective pseudospin-$1/2$ quantum spin-ice model on a microscopic basis\cite{Onoda:10,Onoda:11,Balents:11}. The model also reproduces our polarisation-dependent magnetic neutron-scattering data above $T_{\mathrm{C}}$, showing dipolar spin correlations with a remnant of a pinch-point singularity as a signature of an unusual finite-temperature paramagnetic phase which in practice can be described as a classical magnetic Coulomb liquid\cite{Fennell:09,Morris:09,Isakov:04,Henley:05}.

\section*{Results}

\subsection{Diffuse scattering and pinch-point features above $\mathbf{T_{\mathrm{C}}}$}

Figure~2~\textbf{a} shows the experimental magnetic diffuse neutron-scattering cross-sections collected at 0.3~K in the energy-integrated diffraction mode with an integration over the energy transfer up to 3.64~meV. The background has been subtracted (Supplementary Methods). Here, the incident neutron spins were polarised along the $Z$ direction, perpendicular to the $(hhl)$ plane, allowing us to probe the scattering intensities in the spin-flip (SF) and the non-spin-flip (NSF) channels as well as for the sum of these two channels. 
A broad intensity from $(000)$ through $(111)$ to $(222)$ is identified in the sum (Fig.~2~\textbf{a}, right), as previously reported in unpolarised neutron-scattering experiments\cite{Ross:09}. This feature is dominant in the NSF channel (Fig.~2~\textbf{a}, middle). The $[111]$-rod scattering is accompanied by a branch from (111) to (220) in the SF channel (Fig.~2~\textbf{a}, left), and by another from (222) towards (004) in the NSF channel (Fig.~2~\textbf{a}, middle). The anisotropic nature of the spin correlations is clearly seen around $(111)$ in the SF channel (Fig.~2~\textbf{a}, left), suggesting a remnant of the pinch-point singularity\cite{Henley:05} that has been observed as a sharper singularity in dipolar spin ice~\cite{Fennell:09,Morris:09}. In order to examine this structure in more detail, we have performed similar measurements with a higher resolution at 0.3 and 1.5~K. Figures~2~\textbf{d} and \textbf{e} show the neutron-scattering cross-sections for the SF channel along $[11-1]$ and $[111]$ cuts, respectively, through the $(111)$ point. Whereas the intensity along $[111]$ is featureless and simply increases in magnitude on cooling, the intensity along $[11-1]$ clearly shows a growth of the correlation length. This provides direct evidence of the remnant of the pinch-point singularity.

These polarisation-dependent anisotropic neutron-scattering results are reproduced theoretically as follows. The Yb$^{3+}$ $4f$ magnetic moment at a site $\bm{r}$ is described with the ionic Kramers doublet\cite{Hodges:01} and thus the pseudospin-$1/2$ operator $\hat{\bm{S}}_{\bm{r}}=(\hat{S}^x_{\bm{r}},\hat{S}^y_{\bm{r}},\hat{S}^z_{\bm{r}})$ as
$
\hat{\bm{m}}_{\bm{r}}=\mu_B \left(g_\perp(\hat{S}^x_{\bm{r}}\bm{e}^x_{\bm{r}}+\hat{S}^y_{\bm{r}}\bm{e}^y_{\bm{r}})+g^\parallel \hat{S}^z_{\bm{r}}\bm{e}^z_{\bm{r}}\right)
$
 with the locally defined coordinate frame $(\bm{e}^x_{\bm{r}},\bm{e}^y_{\bm{r}},\bm{e}^z_{\bm{r}})$ (Supplementary Methods), where the $Z$ axis points along the $\langle111\rangle$ direction. The effective Hamiltonian is comprised of the magnetic dipolar interaction,
\begin{equation}
H_D=\frac{\mu_0}{4\pi}\sum_{\langle\bm{r},\bm{r}'\rangle}
\left[\frac{\hat{\bm{m}}_{\bm{r}}\cdot\hat{\bm{m}}_{\bm{r}'}}{({\mit\Delta}r)}
-3\frac{(\hat{\bm{m}}_{\bm{r}}\cdot{\mit\Delta}\bm{r})({\mit\Delta}\bm{r}\cdot\hat{\bm{m}}_{\bm{r}'})}{({\mit\Delta}r)^5}\right],
\label{eq:H_D}
\end{equation}
with ${\mit\Delta}r=|{\mit\Delta}\bm{r}|$, and a nearest-neighbour anisotropic superexchange interaction\cite{Onoda:10,Onoda:11},
\begin{eqnarray}
H_{se}&=&\frac{J}{2}\sum_{\langle\bm{r},\bm{r}'\rangle}^{n.n.}
\left[\hat{S}^z_{\bm{r}}\hat{S}^z_{\bm{r}'}
+\delta\hat{S}^+_{\bm{r}}\hat{S}^-_{\bm{r}'}
+qe^{2i\phi_{\bm{r},\bm{r}'}}\hat{S}^+_{\bm{r}}\hat{S}^+_{\bm{r}'}
+Ke^{i\phi_{\bm{r},\bm{r}'}}(\hat{S}^z_{\bm{r}}\hat{S}^+_{\bm{r}'}+\hat{S}^+_{\bm{r}}\hat{S}^z_{\bm{r}'})\right]+h.c.,
\label{eq:H_se}
\end{eqnarray}
with the bond-dependent phase factor $\phi_{\bm{r},\bm{r}'}$ (Supplementary Methods). Here, $J$ is the coupling constant for the Ising interaction found in spin-ice systems\cite{Bramwell:01,Gardner:04}, while the dimensionless parameters $\delta$/$q$ and $K$ represent the relative coupling constants of the planar quantum exchange interaction that respect/break the U(1) pseudospin symmetry and of another interaction that breaks the $2\pi$-rotation symmetry of the pseudospins about the $\langle111\rangle$ direction.
Figure~2~\textbf{b} shows the magnetic neutron-scattering cross-sections calculated at 0.3~K within the random phase approximation (RPA)\cite{Kao:03}. A set of four adjustable parameters $(J,\delta,q,K)=(0.68~\mathrm{K},-0.8,0.2,-1.0)$ reproduces the experimental profiles shown in Fig.~2~\textbf{a}. (Any change of $\delta$, $q$, or $K$ by $\pm0.1$ gives worse results.) The signs and ratios of the coupling constants obtained from the fitting are comparable to those obtained microscopically with a reasonable ratio $V_{pf\pi}/V_{pf\sigma}=0.2$ of the Slater-Koster parameters\cite{Onoda:11}. The values of $(\delta,q,K)$ are also nearly within the error bars of the recent spin-wave analysis of high-field inelastic neutron-scattering results\cite{Balents:11} (Supplementary Methods). The positive coupling constant $J$ for the Ising interaction favours the 2-in, 2-out spin-ice manifold. (Note that the amplitude of $J$ could be underestimated here because of an overestimation of the onset temperature for the instability in the RPA.) Accordingly, a macroscopic number of states are nearly degenerate, in spite of the additional interactions. In the theoretical calculations, the intensity and the sharpness of a remnant of the pinch-point singularity increase with decreasing temperature in the high-temperature paramagnetic phase (Supplementary Figure~S4), in qualitative agreement with our experimental results. These observations support the scenario that the high-temperature phase of Yb$_2$Ti$_2$O$_7$ in the temperature range $T_{\mathrm{C}}<T<2J$ is well described as a classical magnetic Coulomb liquid which hosts the pinch-point features and monopolar spinons that are deconfined with a finite excitation energy.

\subsection{First-order transition to a ferromagnetic state}

In Fig.~2~\textbf{e} we present the temperature dependence in the sum of the SF and NSF magnetic neutron-scattering cross-sections across the $[111]$ rod through $(1.5, 1.5, 1.5)$. The broad intensity increases gradually on cooling from 0.7 to 0.3~K, but falls abruptly at 0.2~K, signalling a sudden change in the spin correlations. Actually, the neutron-scattering profile at 0.04~K (Supplementary Figure~S3~\textbf{d}) clearly shows that the $[111]$-rod scattering loses intensity. We have measured the $(111)$ Bragg peak intensity as a function of temperature on warming and cooling, with the incident neutron spins polarised along the $X$ direction. Figure~3~\textbf{a} shows the dependence of the sum of the SF and NSF channels for small angular deviations from the wavevector $(111)$ (rocking-curve scans) both above and below $T_{\mathrm{C}}$ with the data collected on warming the sample. By subtracting the 1~K data, the magnetic components can be clearly observed below $T_{\mathrm{C}}$ (Fig.~3~\textbf{b}). The integrated intensities for the SF and NSF channels (Fig.~3~\textbf{c}) and for the sum of the two (Fig.~3~\textbf{d}) show a remarkable hysteresis around $T_{\mathrm{C}}$, indicating a first-order magnetic phase transition. It is possible that the magnetic scattering intensity along the diffusive $[111]$ rod is removed to form the magnetic Bragg peaks below $T_{\mathrm{C}}$. Well above $T_{\mathrm{C}}$ the flipping ratio, (the ratio of the NSF to SF cross-sections), at $(111)$ gradually decreases on cooling and then falls more rapidly to unity around $T_{\mathrm{C}}$ (Fig.~3~\textbf{e}). This indicates that the incident neutron spins are already partially depolarised in the high-temperature phase because of short-range ferromagnetic correlations, and are fully depolarised well below $T_{\mathrm{C}}$, pointing to the emergence of macroscopic ferromagnetic domains. Furthermore, the first-order nature of the transition has also been confirmed by a sharp anomaly in the specific heat of the sample, (Supplementary Figure S1), which closely resembles the earlier result\cite{Blote:69}.

The emergent ferromagnetic order without translational symmetry breaking is also understood theoretically by the mean-field analysis (Supplementary Methods) of the same model at $T=0$ with the values of the dimensionless coupling constants $(\delta,q,K)$ determined above. It is found that the mean-field ground state has a non-coplanar pseudospin order with the Ising component $\langle S^z_{\bm{r}}\rangle\sim0.43$ and the planar component $\sqrt{\langle S^x_{\bm{r}}\rangle^2+\langle S^y_{\bm{r}}\rangle^2}\sim0.25$, as shown in Fig.~1~\textbf{b}. This is translated to a non-coplanar ferromagnetic order with a slight tilting from the $[100]$ direction by $\sim1^\circ$ and a moment amplitude $1.3\mu_B$ (Fig.~1~\textbf{c}), in fair agreement with the previous neutron diffraction results on the same single crystal\cite{Yasui:03}.
In this low-temperature phase, no soft spin excitations exist because of the absence of a continuous symmetry in the original model. This slightly noncoplanar ferromagnetic ground state does not alter with $(\delta,q,K)$ within the accuracy of our fitting procedure.

\section*{Discussion}

From the viewpoint of gauge theory\cite{Hermele:04,Savary:11}, the high-temperature phase of our system is effectively described as a Coulomb phase, where bosonic fractionalised quasi-particles, ``spinons'', carrying magnetic monopole charges are deconfined without a static non-uniform proliferation of monopole charges and host fluctuating U(1) gauge fields responsible for the analogous ``quantum electrodynamics''. These magnetic monopolar spinons can undergo a Bose-Einstein condensation\cite{Banerjee:08} to form the classical ferromagnetic moment\cite{Savary:11} (Fig.~1~\textbf{a}). This pins the global U(1) phase of the spinon fields, and breaks the nontrivial emergent U(1) gauge structure. This condensation of matter (spinon) fields coupled to gauge fields occurs through the Higgs mechanism\cite{Fradkin:79}. Thus, the ferromagnetic state can be viewed as a Higgs phase of magnetic monopoles.  

Our observations support a novel picture that magnetic monopole charges are carried by fractionalised bosonic quasi-particles in the high-temperature magnetic Coulomb liquid. At present, the disappearance of the low-energy gapless excitation spectrum, unique to the Higgs phase, is only suggested by the long magnetic relaxation times seen in this and earlier studies\cite{Yasui:03}, and must be confirmed by future experiments, which could also identify the spectrum of gapped spin wave excitations playing the role of Higgs bosons. It will also be fascinating to search for the quantum phase transition between the magnetic Coulomb phase and the Higgs phase with proper control of the material by pressure or chemical substitution.


\begin{methods}
\subsection{Single-crystal growth}

Single crystals of Yb$_2$Ti$_2$O7, which were also used in the previous published work\cite{Yasui:03}, were prepared by the floating zone method. Stoichiometric quantities of Yb$_2$O$_3$ and TiO$_2$ powder were mixed, pressed into rods, and sintered at $1150^\circ$C for 24 hours. Using these rods, single crystals were grown in air at a rate of 1.5 mm/h. The crystals had a typical diameter of $\sim 6$~mm and a length of 20~mm. Powder X-ray diffraction measurements on a pulverised part of the single crystal showed no appreciable amount of any impurity phase.  

For comparison with other experimental work which reported the absence of long-range magnetic order, the different methods of sample preparation are worth mentioning. Only two reports are available in the literature. According to ref.~\citeonline{Gardner:04}, Gardner \textit{et al.} synthesised polycrystalline samples of Yb$_2$Ti$_2$O$_7$ by firing stoichiometric amounts of Yb$_2$O$_3$ and TiO$_2$ at $1350^\circ$C for several days, with the quality checked by X-ray diffraction. According to ref.~\citeonline{Ross:09}, Ross \textit{et al.} synthesised single crystals of Yb$_2$Ti$_2$O$_7$ using the same floating zone method as our work, but in 4 bars of oxygen and at a faster growth rate of 5 mm/h. Single crystals prepared by Yaouanc \textit{et al.} were also grown by the floating zone method but at an even faster growth rate of 8 mm/h in air with and without a subsequent heat treatment for 24 hours at 1100$^\circ$C under oxygen flow\cite{Yaouanc:11}.

\subsection{Characterisation of different single-crystal samples}

Here, we discuss the controversy over the presence or absence of a first-order phase transition in Yb$_2$Ti$_2$O$_7$. We then demonstrate from measurements of the specific heat and the extended X-ray absorption fine structure (EXAFS) that the quality of the samples significantly alters the low-temperature behaviour of this material. In particular, we show that the long-range ferromagnetic order realised through the first-order phase transition in the single-crystal sample reported in the main text is easily removed by reducing the Yb content within the sample.

First, let us briefly summarise the previous experimental reports that catalogue the strong sample dependence of the physical properties of Yb$_2$Ti$_2$O$_7$. The Curie-Weiss temperature, $\Theta_{\rm CW}$, varies considerably from sample to sample. $\Theta_{\rm CW}\sim0.53$~K for our sample reported on in the main text and in ref.~\citeonline{Yasui:03}, 0.40~K in ref.~\citeonline{Blote:69}, 0.59~K in ref.~\citeonline{Bramwell:00}, and $0.75$~K in ref.~\citeonline{Hodges:01}.  A significant sample dependence is also observed in the low-temperature specific heat in both powder and single-crystal Yb$_2$Ti$_2$O$_7$ samples \cite{Ross:11,Yaouanc:11}. In addition, according to Gardner \textit{et al.}~\cite{Gardner:04}, a magnetic Bragg peak was observed at (111) with polarised neutron-scattering measurements on polycrystalline samples, although neutron spin depolarisation was not observed and static ferromagnetic order was ruled out.

In an attempt to rationalise these conflicting reports, we have studied three single-crystal samples of Yb$_2$Ti$_2$O$_7$ using specific-heat and EXAFS measurements. Sample A was prepared using the method described above and was the sample used for both our polarised neutron-scattering measurements and the previous unpolarised measurements~\cite{Yasui:03}. Sample B was produced from a rod sintered at 1350$^\circ$ rather than 1150$^\circ$ and was grown in air using a faster growth rate of 5 mm/h (cf. ref.~\citeonline{Ross:09} who also used a faster growth rate but grew in 4 bars oxygen). Sample C was grown from another rod which was prepared in the same manner as sample A. However, the molten zone was less stable during the crystal growth.

Supplementary Figure~S1 shows the temperature variation of the specific heat for these three samples. Sample A has a sharp singularity at 0.214~K. This is consistent with the observation made in our neutron-scattering experiments of the first-order phase transition to a ferromagnetic state. Similar sharp anomalies were also observed for polycrystalline samples in both the early and recent measurements~\cite{Blote:69,Ross:11}. However, neutron-scattering experiments have not been reported for powder samples showing such an anomaly. For sample B, the specific-heat anomaly is broadened significantly, signalling the disappearance of the first-order phase transition. The temperature dependence of the specific heat of sample B closely resembles those of the single crystals~\cite{Ross:11,Yaouanc:11} used for unpolarised neutron scattering experiments \cite{Ross:09,Ross:11,Bonville:04}. In sample C, the peak in the heat capacity disappears completely.

The crystal structure of these three samples has been examined by EXAFS at the Yb $L_3$-edge. Supplementary Figure~S2 shows the radial distribution functions of these samples extracted from the Fourier transform of EXAFS oscillations, shown in the upper inset. The oscillations become increasingly damped for sample A (black), B (red) and C (green) respectively, indicating either a higher level of disorder or the appearance of more vacancies for sample A through to sample C. The Fourier transformed radial distribution functions favours the vacancy scenario. Near-edge spectra reveal only the Yb$^{3+}$ charge state (see the middle inset). More oxygen vacancies, particularly at the O(2) site located at the centres of Yb$_4$O tetrahedra, are created than Yb and Ti vacancies in samples B and C. These observations suggest vacancies of all three elements are created cooperatively to maintain charge neutrality. Comparing the EXAFS data with the specific heat curves, one can conclude that the higher the quality of the sample, i.e. the closer the sample is to stoichiometry, the sharper the magnetic transition. In particular, the low-temperature ferromagnetism discussed in the main text is intrinsic to the most ideal single crystals such as sample A.

\subsection{Polarised neutron-scattering experiments}

Measurements were performed at the high-flux polarised diffuse neutron-scattering spectrometer DNS, FRM-II (Garching, Germany). A $^3$He/$^4$He dilution refrigerator insert with an Oxford Instruments cryostat was used for the experimental temperatures from 0.04 to 1 K. A neutron wavelength of 4.74~\AA\ was chosen for all the experiments. The [1-1 0] direction of the crystals was aligned perpendicular to the horizontal scattering plane so that the $(h h l)$ reciprocal plane can be mapped out by rotating the sample. The neutron polarisation at the sample position was aligned along the [1-10] direction of the sample, i.e. the $Z$-direction of the chosen experimental coordinate system ($Z$-direction polarised neutron scattering). Within this setup, the SF and NSF scattering cross-sections are given by
\begin{eqnarray}
  \left(\frac{d\sigma}{d\Omega}\right)^{Z-\mathrm{SF}}&=&M_{\perp Y}^* M_{\perp Y}+\frac{2}{3}I_{SI},
  \label{eq:SF:Z}\\
  \left(\frac{d\sigma}{d\Omega}\right)^{Z-\mathrm{NSF}}&=&M_{\perp Z}^*M_{\perp Z}+N^*N+\frac{1}{3}I_{SI},
  \label{eq:NSF:Z}
\end{eqnarray}
respectively, where 
$M_{\perp Y}^*M_{\perp Y}$ and $M_{\perp Z}^*M_{\perp Z}$ are the components of the magnetic scattering cross-section in and out of the $(h h l)$ scattering plane respectively, with $Y$ being perpendicular to the scattering wavevector $\bm{Q}$. $I_{SI}$ is the total nuclear spin incoherent scattering cross-section, and $N^*N$ is the nuclear coherent scattering cross-section. In general, the magnetic scattering cross-sections $M_{\perp Y}^*M_{\perp Y}$ and $M_{\perp Z}^*M_{\perp Z}$ are not identical. Therefore, they can provide information on the anisotropy of the magnetic correlations present in the system. In order to confirm the observed magnetic scattering, $X$-direction polarised neutron scattering has also been carried out. In this polarised neutron-scattering setup, the neutron polarisation is parallel to the scattering vector $\bm{Q}$, and can be used to separate the magnetic scattering contribution to be mapped in the SF channel, and nuclear contribution in the NSF channel;
\begin{eqnarray}
  \left(\frac{d\sigma}{d\Omega}\right)^{X-\mathrm{SF}}&=&M_{\perp Y}^* M_{\perp Y}+M_{\perp Z}^*M_{\perp Z}+\frac{2}{3}I_{SI},
  \label{eq:SF:X}\\
  \left(\frac{d\sigma}{d\Omega}\right)^{X-\mathrm{NSF}}&=&N^*N+\frac{1}{3}I_{SI}.
  \label{eq:NSF:X}
\end{eqnarray}
 For details, see refs~\citeonline{PNS1,PNS2}.

\end{methods}

\begin{addendum}

 \item We thank J. S. Gardner and B. D. Gaulin for discussions. We thank Harald Schneider for technical supports in the neutron-scattering experiments. We acknowledge Chih-Wen Pao and other staff of BL07A of the NSRRC who helped perform the EXAFS measurements. 
This work is partially supported
by National Science Council, Taiwan, under Grants No. NSC 96-2739-M-213-001, NSC 99-2112-M-007-020 (L.J.C.), NSC 100-2112-M-002 -013 -MY3, NSC-99-2120-M-002-005 (Y.J.K.), by NTU under Grant No. 10R80909-4 (Y.J.K.), and by Grants-in-Aid for Scientific Research under Grant No. 21740275 and No. 24740253 from Japan Society for the Promotion of Science (S.O.) and under Grant No. 19052006 (S.O.) and No. 19052004 (K.K.) from the Ministry of Education, Culture, Sports, Science and Technology, Japan.

 \item[Author Contributions] L.J.C. performed the neutron-scattering experiments and wrote the manuscript; S.O. and Y.J.K. performed the theoretical calculations and wrote the manuscript; Y.S. designed and performed the neutron-scattering experiments; Y.Y. grew the single-crystal sample; K.K. and M.R.L. supervised the neutron-scattering experiments. M.R.L. carried out the heat capacity measurements. K.D.T. performed the EXAFS studies. All the authors discussed the results and the manuscript.

\item[Subject terms]
condensed matter physics, materials physics, magnetic materials and devices
 \item[Supplementary Information] accompanies this paper at http://www.nature.com/naturecommunications

 \item[Competing financial interests:] The authors declare that they have no competing financial interests.

 \item[Correspondence] Correspondence and requests for materials should be addressed to L.J.C. \\
(email: ljchang@mail.ncku.edu.tw) 
or S.O. (email: s.onoda@riken.jp).
\end{addendum}

\newpage
\begin{figure}
\caption{\textbf{Schematic phase diagram as a function of temperature $T$ and the relative strength $\delta$ of the U(1)-symmetric planar exchange interaction and the Ising exchange, and the hypothetical low-temperature ordered structures of Yb$_2$Ti$_2$O$_7$.}
\textbf{a}, Schematic phase diagram showing the first-order Higgs transition between a Coulomb liquid phase and a Higgs phase of magnetic monopoles. The other two model parameters $q$ and $K$ in Eq.~(\ref{eq:H_se}) are assumed to be negligibly small for dipolar spin ice (Dy/Ho)$_2$Ti$_2$O$_7$, while they are finite for Yb$_2$Ti$_2$O$_7$ as obtained in the present work. Monopoles (blue balls) and anti-monopoles (red balls) are illustrated for both the phases.
In the magnetic Coulomb liquid phase (yellow), magnetic monopoles are carried by pseudospin-$1/2$ fractionalised gapped spinon excitations out of quasi-degenerate spin-ice manifold, obeying a Coulombic law. In the Higgs phase (cyan), monopolar spinons are condensed to form local magnetic dipole moments (arrows) showing ferromagnetic long-range order.
\textbf{b}, \textbf{c}, Hypothetical ferromagnetically ordered structures of the pseudospins (\textbf{b}) and the magnetic moments (\textbf{c}) in the low-temperature Higgs phase. The finite planar components of the pseudospins are ascribed to a condensation of monopolar spinons in the U(1) gauge theory\cite{Hermele:04,Savary:11}.}
\end{figure}

\newpage
\begin{figure}
\caption{\textbf{Magnetic diffuse neutron-scattering profiles in the $(hhl)$ plane with the incident neutron spins polarised along the $Z$ direction.}
\textbf{a}, Experimentally observed profiles for the SF (left), NSF (middle) and total (right) magnetic diffuse neutron scattering cross-sections above $T_{\mathrm{C}}$ at $T=0.3$~K. The background levels for the diffuse scattering have been subtracted by making use of the data measured with the incident neutron beam polarised along the $X$ direction parallel to the scattering wavevector $\bm{q}$ (see Supplementary Methods and Supplementary Figure~S3). White circles mark the Bragg spots where the background subtraction scheme is less reliable. The measurements were performed in the energy-integrated diffraction mode up to an energy of energy 3.64~meV.
\textbf{b}, Theoretically calculated magnetic profiles for SF (left), NSF (middle) and total (right) magnetic diffuse neutron scattering cross-sections at $T=0.3$~K. The profiles were obtained as the equal-time correlators in the random phase approximation (Supplementary Methods) to the pseudospin-$1/2$ model, $H_D+H_{se}$. The form factor is not included.
\textbf{c} and \textbf{d}, Neutron-scattering intensity for the SF channel along $[11-1]$ and $[111]$ cuts, respectively, across the remnant of the pinch point $(111)$ at $T=0.3$ and 1.5~K. The data were measured with a higher resolution in reciprocal space than the full $(hhl)$ profiles. One or two points marking a high level of intensity at and close to the $(111)$ point correspond to the nuclear Bragg peak. The lines are a guide to the eye.
\textbf{e}, Temperature dependence of the total cross-section along $[11-1]$ cut through $(1.5,1.5,1.5)$, as indicated in the inset.
}
\end{figure}

\newpage
\begin{figure}
\caption{\textbf{Hysteresis in the Bragg peak intensity and flipping ratio at $(111)$ on cooling and warming across $T_{\mathrm{C}}$, measured in the $X$-direction polarised neutron-scattering experiments.} 
In order to ensure that the system was in thermal equilibrium, measurements well below $T_{\rm{C}}$ were performed after waiting longer than the magnetic relaxation time, which is of the order of 2 hours at 0.03~K~(ref.~\citeonline{Yasui:03}).
{\bf a}, The rocking-curve scans of the $(111)$ Bragg peak for the sum of the SF and NSF channels measured while warming the sample. The horizontal axis represents the deviation of the wavevector from $(111)$ with their amplitude being unchanged. The lines are a guide to the eye. All the peaks are instrument resolution limited. 
{\bf b}, The magnetic contribution to the $(111)$ Bragg peak is obtained by subtracting the data measured at 1~K, which is well above $T_{\mathrm{C}}$, in the warming sequence.
{\bf c}, The growth of the integrated intensity for the (111) Bragg peak begins abruptly below $T_{\mathrm{C}}$ with a discernible hysteresis in both the SF and the NSF channels.
{\bf d}, The sum of the SF and the NSF $(111)$ Bragg peak intensities versus temperature showing a hysteresis with temperature. The sum increases by almost $6\%$ below $T_{\mathrm{C}}$ in agreement with the previous measurements\cite{Yasui:03}.
{\bf e}, The flipping ratio of the neutron spins at the $(111)$ Bragg peak as a function of temperature. This ratio falls off steeply to unity below $T_{\mathrm{C}}$ with a clear hysteresis.
The data were collected by first warming and then cooling. This sequence is indicated by the arrows in \textbf{c}, \textbf{d} and \textbf{e}.}
\end{figure}

\newpage
\begin{figure}
\begin{center}
\includegraphics[width=\textwidth]{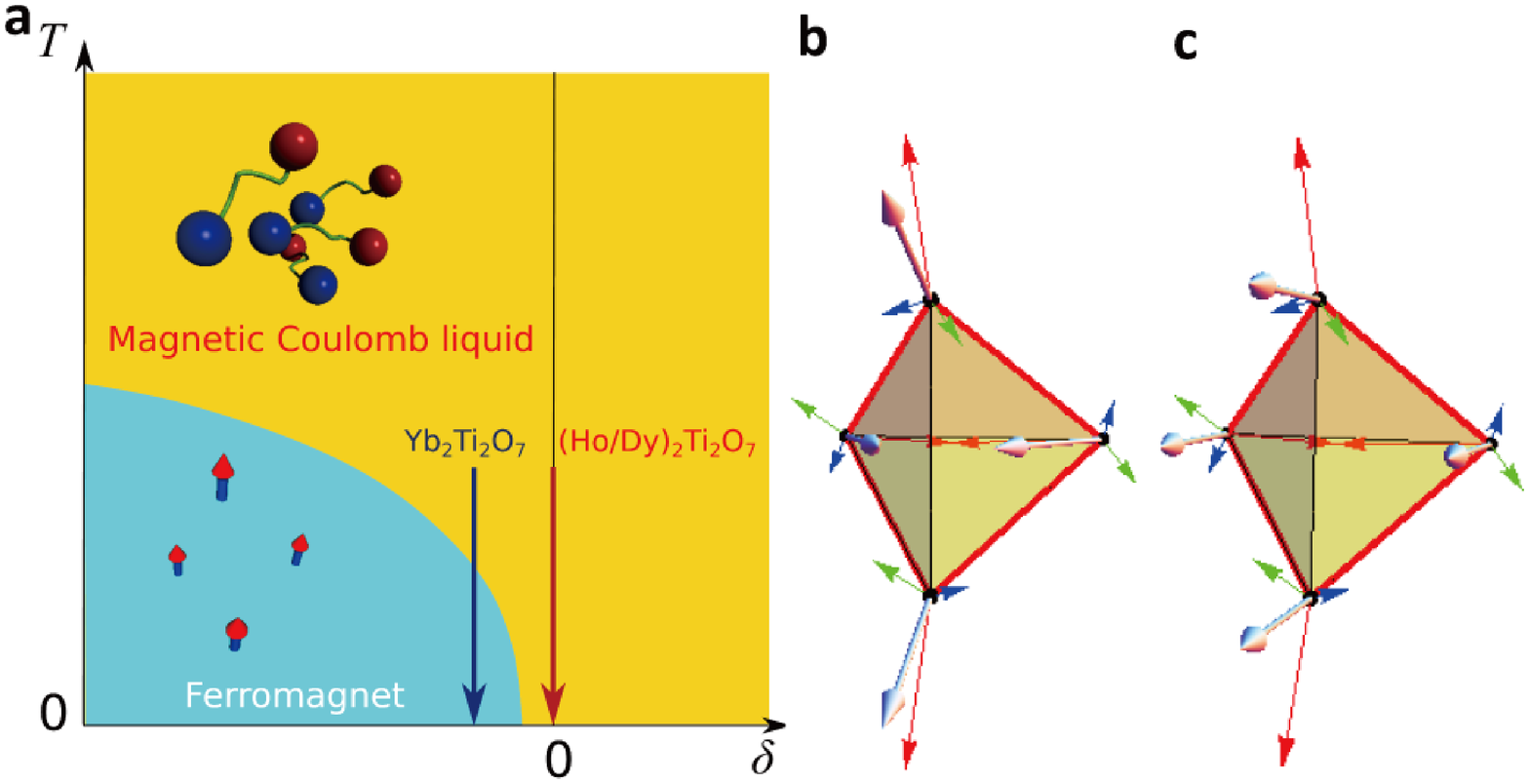}
\end{center}
{\bf Figure 1.}
\end{figure}
\newpage
\begin{figure}
\begin{center}
\includegraphics[width=\textwidth]{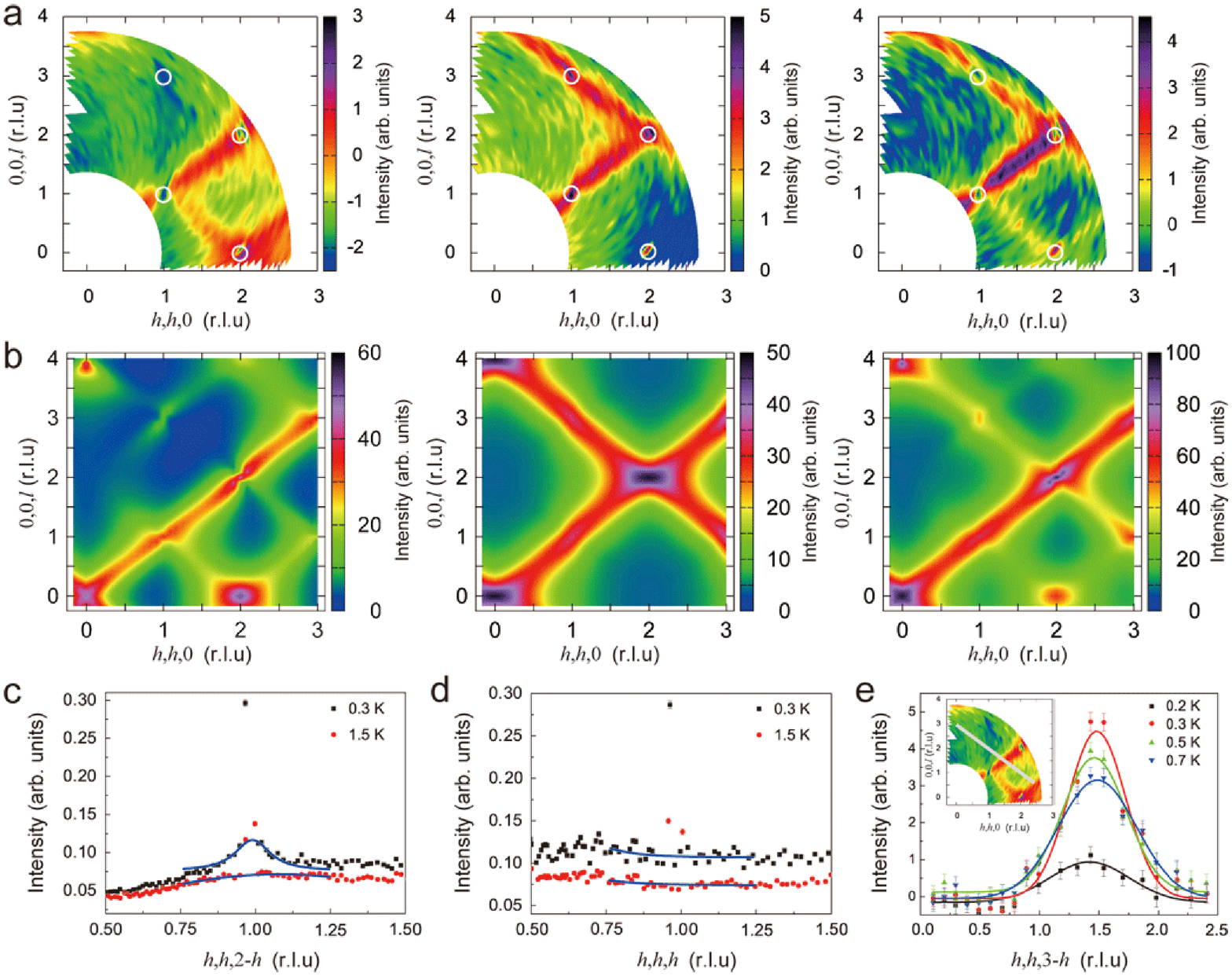}
\end{center}
{\bf Figure 2.}
\end{figure}
\newpage
\begin{figure}
\begin{center}
\includegraphics[width=\textwidth]{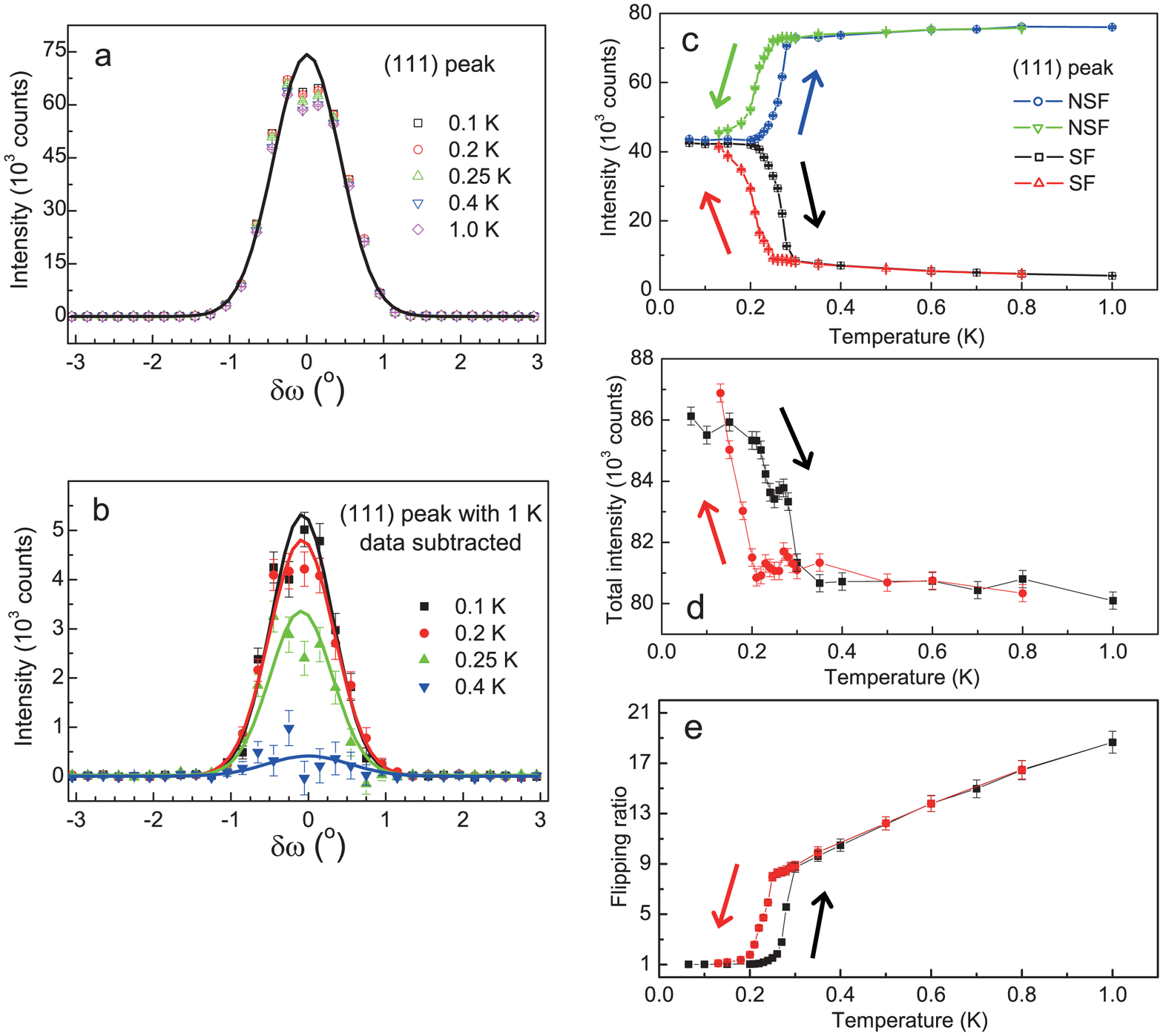}
\end{center}
{\bf Figure 3.}
\end{figure}

\end{document}


\maketitle

\newpage
\section*{Supplementary Figure 1}
\begin{figure}[H]
\begin{center}
\includegraphics[width=0.8\textwidth]{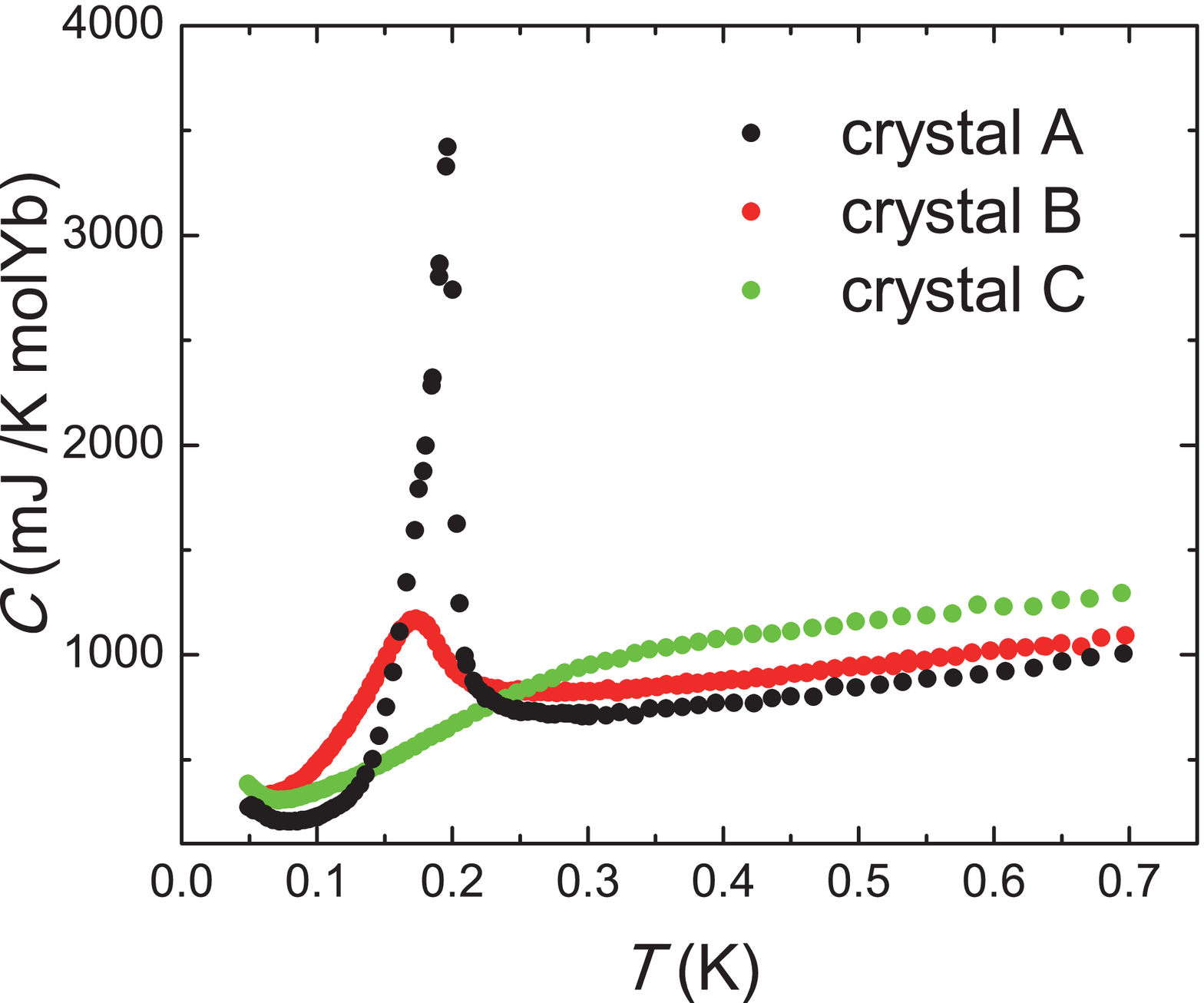}
\end{center}
\caption{\textbf{Low-temperature specific heat versus temperature for three single crystals of Yb$_2$Ti$_2$O$_7$.} Sample A was used for the polarised neutron-scattering experiments reported in the main text. The heat capacity measurements were carried out in a Quantum Design Physical Property Measurement System (PPMS) using a pulse-relaxation calorimetry technique.}
\label{fig:Cv}
\end{figure}

\newpage
\section*{Supplementary Figure 2}
\begin{figure}[H]
\begin{center}
\includegraphics[width=0.6\textwidth]{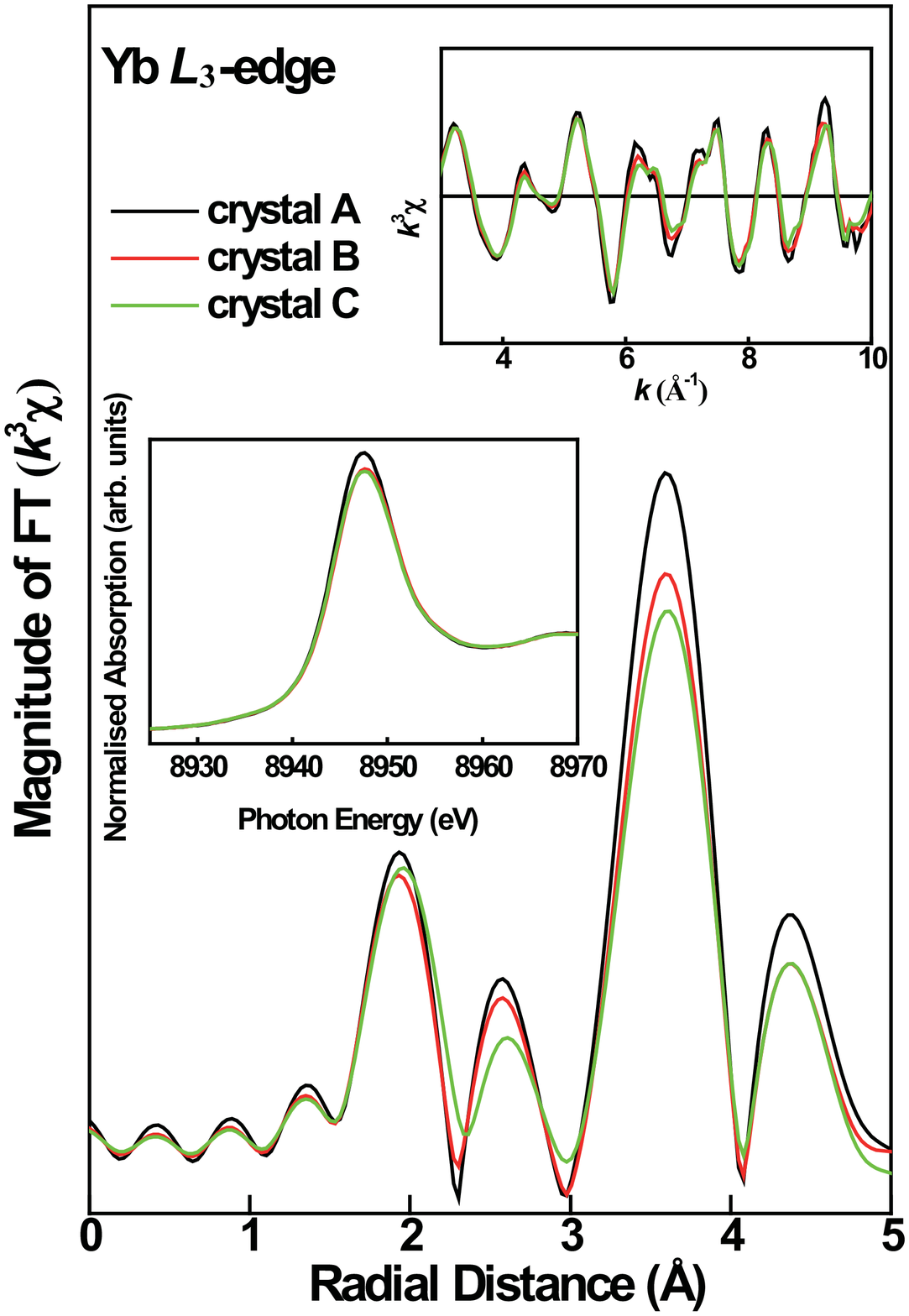}
\end{center}
\caption{\textbf{EXAFS radial distribution functions at the Yb $L_3$-edge from the Fourier transform of raw spectra in the upper inset.} The near edge x-ray absorption spectra (XAS), shown in the middle inset, reveal only the Yb$^{3+}$ peak at 8947~eV. The data are presented for crystals A (black), B (red) and C (green).}
\label{fig:EXAFS}
\end{figure}

\newpage
\section*{Supplementary Figure 3}
\begin{figure}[H]
\begin{center}
\includegraphics[width=0.66\textwidth]{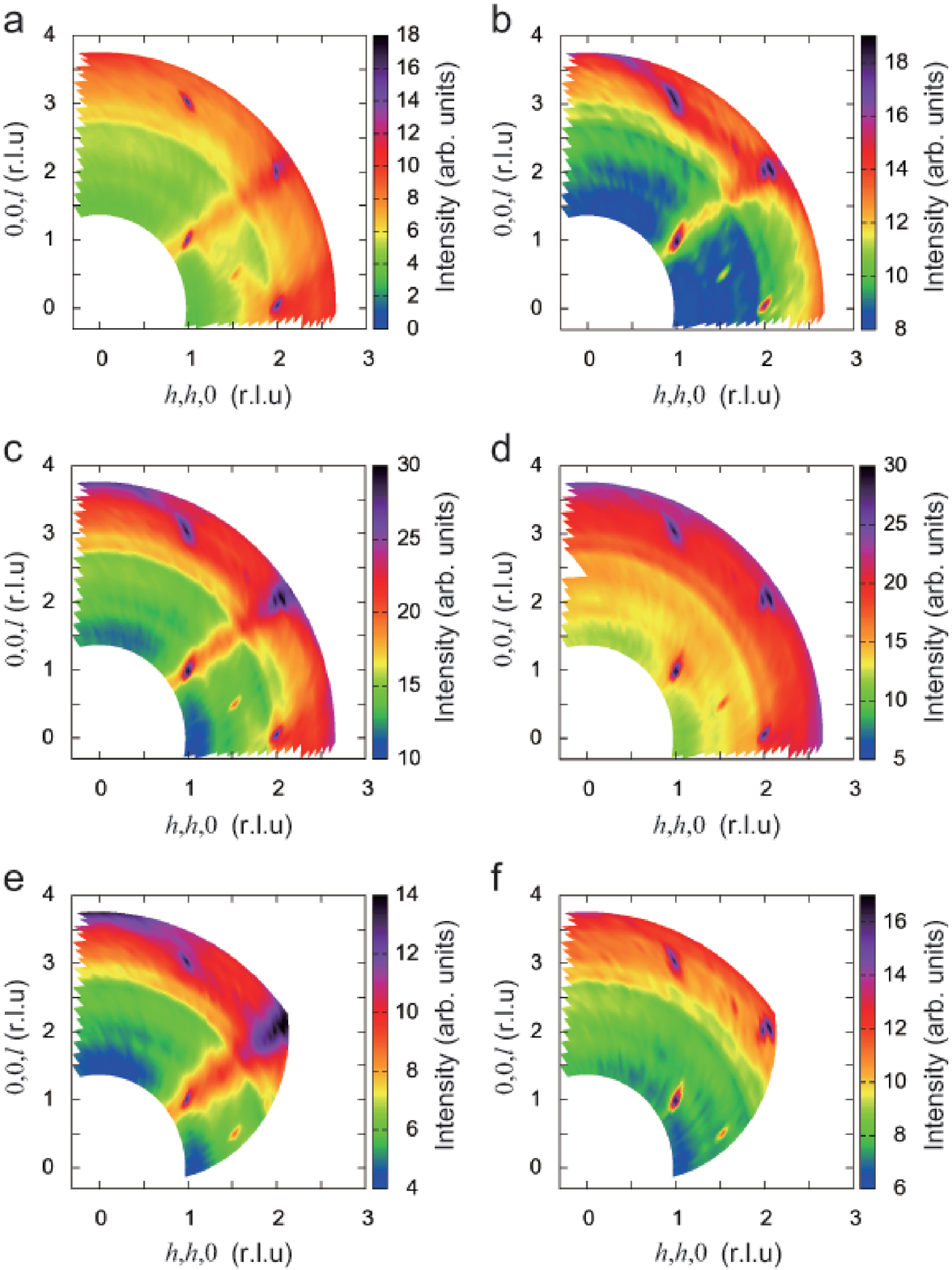}
\end{center}
\caption{\textbf{Raw polarised neutron-scattering cross-sections taken in the energy-integrated diffraction mode at DNS.} \textbf{a}, \textbf{b}, and \textbf{c}, The SF, NSF, and total cross-sections, respectively, in the $Z$-direction polarised neutron-scattering measurements at $T=0.3$~K. \textbf{d}, The total cross-section in the $Z$-direction polarised neutron-scattering measurements at $T_0=0.04$~K. \textbf{e} and \textbf{f}, The SF and NSF cross-sections, respectively, in the $X$-direction polarised neutron-scattering measurements at $T_X=0.7$~K. Sharp peaks at reciprocal lattice vectors $(111)$, $(002)$, $(220)$ and so on are dominated by the nuclear contributions. The weak peaks at $(1.5, 1.5, 0.5)$ and $(112)$ are attributed to contamination from higher order scattering and are not intrinsic.}
\label{fig:ZX}
\end{figure}

\newpage
\section*{Supplementary Figure 4}
\begin{figure}[H]
\begin{center}
\includegraphics[width=0.95\textwidth]{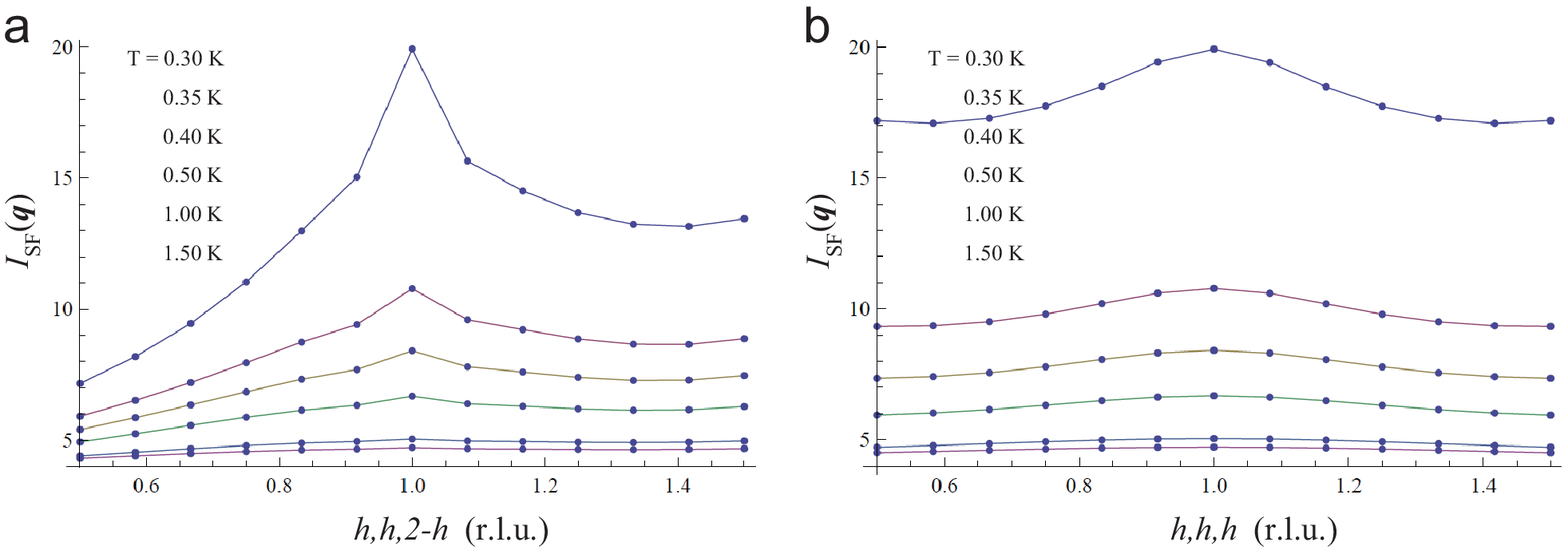}
\end{center}
\caption{\textbf{Temperature dependence of the theoretical calculated equal-time correlator corresponding to the energy-integrated $Z$-polarised neutron-scattering cross-section in the SF channel.} \textbf{a}, The intensity along the $[11-1]$ cut through the $(111)$ point. \textbf{b}, The intensity along the $[111]$ cut through the $(111)$ point. }
\label{fig:pinch-point}
\end{figure}

\newpage
\section*{Supplementary Methods}

\subsection*{Background subtraction}

Here, we explain how we derive the $Z$-direction polarised diffuse magnetic scattering profiles at 0.3~K shown in Figure~2~\textbf{a} of the main text from the raw data shown in Supplementary Figure~\ref{fig:ZX}. We estimate and subtract the background levels for the SF and NSF channels as follows.

Because of the formation of a long-range ferromagnetic order and the loss of the $[111]$ rod diffuse scattering intensity well below $T_{\mathrm{C}}$~K, the sum of the SF and NSF channels at $T_0=0.04$~K can be roughly regarded as a temperature-independent background due to the nuclear coherent and incoherent scattering cross-sections;
\begin{equation}
  \left(\frac{d\sigma}{d\Omega}\right)^{\mathrm{Total\  bg}}
  =
  N^*N+I_{SI}
  \approx
  \left(\frac{d\sigma}{d\Omega}\right)^{\mathrm{Total}}_{T_0},
  \label{eq:total:bg}
\end{equation}
except around the Bragg peak positions where we suffer from a slight overestimate due to the magnetic Bragg-peak intensities.
On the other hand, the SF and NSF cross-section profiles obtained at 0.04~K cannot be used as background levels, because the incident neutron spins are fully depolarised at 0.04~K, while they are only partially depolarised at 0.3~K, as is clear from Figure~3~\textbf{e} of the main text. 

Note that the spin flipping ratio is defined as the ratio of the NSF to SF scattering intensity at a Bragg-peak position $(111)$, where nuclear coherent scattering cross-section $N^*N$, which appears only in the NSF channel according to 
Eqs.~(5) and (6) 
as well as 
Eqs.~(3) and (4), dominates over the other contributions. Then, it is reasonable to assume that the SF and NSF cross-sections observed at a temperature $T$ with the $Z$- and $X$-direction polarised neutron-scattering measurements take the forms
\begin{eqnarray}
  \left(\frac{d\sigma}{d\Omega}\right)^{Z-\mathrm{SF}}_{T}
  &\approx&
  \frac{\left(M_{\perp Y}^*M_{\perp Y}\right)_T+\frac{2}{3}I_{SI}
    +r_T^{-1}\left[\left(M_{\perp Z}^*M_{\perp Z}\right)_T+N^*N+\frac{1}{3}I_{SI}\right]}{1+r_T^{-1}},
  \label{eq:ZSF}\\
  \left(\frac{d\sigma}{d\Omega}\right)^{Z-\mathrm{NSF}}_{T}
  &\approx&
  \frac{r_T^{-1}\left[\left(M_{\perp Y}^*M_{\perp Y}\right)_T+\frac{2}{3}I_{SI}\right]+\left(M_{\perp Z}^*M_{\perp Z}\right)_T+N^*N+\frac{1}{3}I_{SI}}{1+r_T^{-1}},
  \label{eq:ZNSF}\\
  \left(\frac{d\sigma}{d\Omega}\right)^{X-\mathrm{SF}}_{T}
  &\approx&
  \frac{\left(M_{\perp Y}^*M_{\perp Y}+M_{\perp Z}^*M_{\perp Z}\right)_T+\frac{2}{3}I_{SI}+r_T^{-1}\left[N^*N+\frac{1}{3}I_{SI}\right]}{1+r_T^{-1}},
  \label{eq:XSF}\\
  \left(\frac{d\sigma}{d\Omega}\right)^{X-\mathrm{NSF}}_{T}
  &\approx&
  \frac{r_T^{-1}\left[\left(M_{\perp Y}^*M_{\perp Y}+M_{\perp Z}^*M_{\perp Z}\right)_T+\frac{2}{3}I_{SI}\right]+N^*N+\frac{1}{3}I_{SI}}{1+r_T^{-1}}.
  \label{eq:XNSF}
\end{eqnarray}
Using Eqs.~(\ref{eq:XSF}) and (\ref{eq:XNSF}), the magnetic cross-section can be eliminated and the temperature-independent purely $X$-NSF background cross-section is given by
\begin{equation}
  N^*N+\frac{1}{3}I_{SI}
  =
  \frac{\left(\frac{d\sigma}{d\Omega}\right)^{X-\mathrm{NSF}}_{T_X}
  -r_{T_X}^{-1}\left(\frac{d\sigma}{d\Omega}\right)^{X-\mathrm{SF}}_{T_X}}
  {1-r_{T_X}^{-1}},
  \label{eq:NSF:bg}
\end{equation}
at a temperature $T_X$. For a higher accuracy, it is better to take the data with a larger spin flipping ratio $r_{T_X}$, and thus we take $T_X=0.7$~K where $r_{T_X}\sim14.97$ (Figure~3~\textbf{e} of the main text). The raw data for the $X$-direction polarised elastic diffuse neutron-scattering experiments are shown in Supplementary Figures~\ref{fig:ZX}~\textbf{e} and \ref{fig:ZX}~\textbf{f}.
In fact, the $\bm{q}$-space region where the $X$-direction polarised experiments have been performed does not cover all the reciprocal space required for this analysis. Nevertheless, the profile for Eq.~(\ref{eq:NSF:bg}) obtained from Supplementary Figures~\ref{fig:ZX}~\textbf{e} and \ref{fig:ZX}~\textbf{f} at $T_X=0.7$~K is reproduced to an accuracy of $10\%$, except at the Bragg-peak positions, by multiplying the profile for $\left(\frac{d\sigma}{d\Omega}\right)^{\mathrm{total}}_{T_0}$ at $T_0=0.04$~K shown in Supplementary Figure~\ref{fig:ZX}~\textbf{d} by an overall scale factor 0.6. This allows us to simulate the NSF background profile for Eq.~(\ref{eq:NSF:bg}) in the whole required area of the $(hhl)$ plane.

Now, we can express the background cross-sections for the SF and NSF channels in the $Z$-direction polarised neutron-scattering experiments as
\begin{eqnarray}
  \left(\frac{d\sigma}{d\Omega}\right)^{Z-\mathrm{SF}\ \mathrm{bg}}_T
  &=&
  \frac{\left[N^*N+I_{SI}\right]-(1-r_T^{-1})\left[N^*N+\frac{1}{3}I_{SI}\right]}{1+r_T^{-1}},
  \label{eq:SF:Z:bg}\\
  \left(\frac{d\sigma}{d\Omega}\right)^{Z-\mathrm{NSF}\ \mathrm{bg}}_T
  &=&
  \frac{r_T^{-1}\left[N^*N+I_{SI}\right]+(1-r_T^{-1})\left[N^*N+\frac{1}{3}I_{SI}\right]}{1+r_T^{-1}},
  \label{eq:NSF:Z:bg}
\end{eqnarray}
where $\left[N^*N+I_{SI}\right]$ and $\left[N^*N+\frac{1}{3}I_{SI}\right]$ are given by Eqs.~(\ref{eq:total:bg}) and (\ref{eq:NSF:bg}), respectively. The spin flipping ratio at $T=0.3$~K is given by $r_{T}=8.7$ (Figure~3~\textbf{e} of the main text).

In principle, we should also be able to eliminate a small ($11.5\%$) mixing of $M_{\perp Z}^*M_{\perp Z}$ in the SF channel and of $M_{\perp Y}^*M_{\perp Y}$ in the NSF channel at 0.3~K. However, this effect is found to be small compared with the accuracy of the background subtraction.

\subsection*{Mean-field theory and random phase approximations}

The classical mean-field (MF) Hamiltonian is given by
\begin{equation}
H_{MF}
=\frac{1}{2}\sum_{\bm{q}}\sum_{\mu,\nu=x,y,z}\sum_{i,j=0}^3
\langle2S_{-\bm{q},i}^\mu\rangle
h^{\mu\nu}_{\bm{q},i,j}
\langle2S_{\bm{q},j}^\mu\rangle
\end{equation}
where $h^{\mu\nu}_{\bm{q},i,j}$ is obtained from the Fourier transform of the full Hamiltonian $(H_D+H_{se})$ given in the main text$^{16,19}$,
and $i$ and $j$ denote indices for the Yb sites at $\bm{R}+\bm{a}_i$ ($i=0,\cdots,3$) with fcc lattice vectors $\bm{R}$ and $\bm{a}_0=(0,0,0)$, $\bm{a}_1=a(0,\frac{1}{4},\frac{1}{4})$, $\bm{a}_2=a(\frac{1}{4},0,\frac{1}{4})$, and $\bm{a}_3=a(\frac{1}{4},\frac{1}{4},0)$, and the cubic lattice constant $a$. The local coordinate frames are taken as
\begin{eqnarray}
  \bm{e}^x_0=-\frac{1}{\sqrt{6}}\left(1,1,-2\right),
  \ \ \
  \bm{e}^y_0=-\frac{1}{\sqrt{2}}\left(-1,1,0\right),
  \ \ \ 
  \bm{e}^z_0=\frac{1}{\sqrt{3}}(1,1,1),
  \label{eq:xyz0}\\
  \bm{e}^x_1=-\frac{1}{\sqrt{6}}\left(1,-1,2\right),
  \ \ \ 
  \bm{e}^y_1=-\frac{1}{\sqrt{2}}\left(-1,-1,0\right),
  \ \ \
  \bm{e}^z_1=\frac{1}{\sqrt{3}}(1,-1,-1),
  \label{eq:xyz1}\\
  \bm{e}^x_2=-\frac{1}{\sqrt{6}}\left(-1,1,2\right),
  \ \ \ 
  \bm{e}^y_2=-\frac{1}{\sqrt{2}}\left(1,1,0\right),
  \ \ \ 
  \bm{e}^z_2=\frac{1}{\sqrt{3}}(-1,1,-1),
  \label{eq:xyz2}\\
  \bm{e}^x_3=-\frac{1}{\sqrt{6}}\left(-1,-1,-2\right),
  \ \ \ 
  \bm{e}^y_3=-\frac{1}{\sqrt{2}}\left(1,-1,0\right),
  \ \ \ 
  \bm{e}^z_3=\frac{1}{\sqrt{3}}(-1,-1,1),
  \label{eq:xyz3}
\end{eqnarray}
to maximally exploit the symmetry of the system. In particular, the $z$ axis points to the $\langle111\rangle$ direction of the Ising magnetic moment. This in turn introduces the following simple bond-dependent phases $\phi_{\bm{r},\bm{r}'}=\phi_{ij}=\phi_{ji}$ in the two superexchange terms proportional to $q$ and $K$ which break the planar U(1) pseudospin symmetry;
\begin{eqnarray}
\phi_{01}=\phi_{23}=-\frac{2\pi}{3},~~\phi_{02}=\phi_{31}=\frac{2\pi}{3},~~\phi_{03}=\phi_{12}=0,
\end{eqnarray}
where $(i,j)$ is a pair of sublattice indices for $\bm{r}$ and $\bm{r}'$.

The Ewald summation technique was employed to calculate the long-range magnetic dipolar interaction. Then, the eigenvalue problem is solved for each $\bm{q}$ as 
\begin{equation}
\sum_{\nu}h^{\mu\nu}_{\bm{q},i,j}\phi^\nu_{\bm{q},n,j}=\varepsilon_{\bm{q},n}\phi^\mu_{\bm{q},n,i},
\end{equation}
where $n$ labels the twelve eigenvalues/eigenstates. 
The magnetic neutron-scattering cross-sections at a temperature $T$ are then calculated with the random phase approximation (RPA)$^{34,44}$
as
\begin{eqnarray}
I(\bm{q})=\sum_{n}\sum_{i,j=1}^4f^{\mu\nu}_{\bm{q},i,j}
\frac{\phi^\mu_{-\bm{q},n,i}\phi^\nu_{\bm{q},n,j}}{3+\epsilon_{\bm{q},n}/T}.
\end{eqnarray}
where 
\begin{equation}
f^{\mu\nu}_{\bm{q},i,j}=\frac{1}{4|\bm{q}|^4}\left[\bm{Z}\cdot(\bm{q}\times(g_\mu\bm{e}_i^\mu\times\bm{q}))\right] \left[\bm{Z}\cdot(\bm{q}\times(g_\nu\bm{e}_j^\nu\times\bm{q}))\right],
\end{equation}
with $\bm{Z}=(1,-1,0)/\sqrt{2}$ for the NSF channel and 
\begin{equation}
f^{\mu\nu}_{\bm{q},i,j}=\frac{1}{4|\bm{q}|^4}\left[\bm{q}\times(g_\mu\bm{e}_i^\mu\times\bm{q})\right] \cdot \left[\bm{q}\times(g_\nu\bm{e}_j^\nu\times\bm{q})\right]
\end{equation}
for the total. The SF channel is obtained as the difference between the above two.

In our case of $J>0$, while $\delta>0$ leads to profiles analogous to the nearest-neighbour Ising exchange or dipolar spin ice$^{10}$,
a choice of $\delta<0$ produces various profiles depending on the other relative coupling constants $q$ and $K$. The best fit to our experimentally obtained and analysed magnetic scattering profiles at 0.3~K, shown in Figure~2~\textbf{a} of the main text, was obtained with $(J,\delta,q,K)=(0.68~\mathrm{K},-0.8,0.2,-1.0)$, as shown in Figure~2~\textbf{b} of the main text. This set of parameter values is compared to the result from the spin-wave fitting$^{20}$,
which is translated into $J=2.0\pm0.5$~K, $\delta=-0.6\pm0.3$, $q=0.6\pm0.3$ and $K=-1.6\pm\sim0.4$ in the present notation. Our exchange energy scale is smaller by a factor of 2-3. The discrepancy in $J$ may be partly attributed to an overestimate of the temperature at which an RPA instability occurs, $\sim0.29$~K. This gives $\theta_{\rm CW}\sim0.29$~K. Scaling the energy scale by a factor of 2 reproduces the experimental observation $\theta_{\rm CW}\sim0.53$~K~(ref.~24).
$\delta$ and $q$ are nearly within error bars of the analysis given in ref.~20.
However, we note that if we take $|\delta|$ to be comparable to $q$ and/or take a much larger value of $|K|$, e.g., $K=-1.2$, following ref.~20,
the intensity at the scattering branch from (111) to (220) in the SF channel is significantly suppressed, in disagreement with our experiments.
The nearest-neighbour coupling constants from $H_D$ are of the order of 0.01~K, namely, one or two orders of magnitude smaller than the superexchange couplings of our choice, though they give rise to $\sim-0.2$~K in the energy eigenvalue of the mean-field Hamiltonian. This might cause other minor differences.
 As also noted in ref.~20,
the parameters $\delta$, $q$, and $K$ we have obtained are quite different from the estimates by Thompson \textit{et al.}\cite{Thompson:10} who analysed the Hamiltonian including all the crystal field states but only the same number of coupling constants without any multipolar interactions.

In Supplementary Figure~\ref{fig:pinch-point}~\textbf{a} and \textbf{b}, we show the temperature dependence of the calculated neutron-scattering intensity in the SF channel (without taking account of the form factor) along the $[11-1]$ and $[111]$ cuts through the $(111)$ point, respectively. It is clear that the intensity along the $[11-1]$ cut is not only enhanced but also sharpened with decreasing temperature down to 0.3~K, while that along the $[111]$ cut remains almost featureless even at 0.3~K. This is consistent with the picture that the high-temperature phase of this model is described as a Coulomb phase characterised by a remnant of a pinch-point singularity.

The mean-field ground state was obtained from the same scheme as employed above. We searched for the lowest-energy states among the classical energies per site,
\begin{equation}
E_{cl,\bm{q},n}=\frac{\varepsilon_{\bm{q},n}}{Max_i\sum_\mu|\phi^\mu_{\bm{q},n,i}|^2}.
\end{equation}
Then, the MF ground state is found by approaching the (000) point from the $[100]$ direction, resulting in the noncoplanar pseudospin structure and the nearly collinear ferromagnetic structures shown in Figures~1~\textbf{b} and \textbf{c} of the main text, respectively.

Note that the possibility of an ordered spin-ice state$^{10,45,46}$
is ruled out, since it should create magnetic Bragg peaks at (100), in contradiction to our findings and cannot explain the neutron depolarisation due to the macroscopic magnetic moment.